\documentclass[journal]{IEEEtran}
\usepackage{booktabs}
\usepackage{cite}
\usepackage{amsmath,amssymb,amsfonts}
\usepackage{graphicx}
\usepackage{xcolor}
\usepackage{bm}
\usepackage{multirow}
\usepackage[hidelinks]{hyperref}
\usepackage{xurl}

\graphicspath{{./}{arxiv/}}

\newcommand{\wn}{\omega_{\mathrm n}}
\newcommand{\Da}{D_{\mathrm{act}}}
\newcommand{\Ja}{J_{\mathrm{act}}}

\newcommand{\Je}{J_{\mathrm{eff}}}
\newcommand{\Js}{J_{\mathrm{sup}}}
\newcommand{\dd}{\mathrm{d}}
\DeclareMathOperator{\sinc}{sinc}

\title{Beyond Scalar Inertia: Frequency-Response-Based Assessment of Finite-Duration Inertial Support in Grid-Forming Converters}

\author{
Ziqi Zhou,~\IEEEmembership{Member,~IEEE},
Mahdieh S. Sadabadi,~\IEEEmembership{Senior Member,~IEEE},
Sante Pugliese,~\IEEEmembership{Member,~IEEE},
Xiaoqiang Li,~\IEEEmembership{Member,~IEEE},
Xiaojie Wu,~\IEEEmembership{Member,~IEEE},
and Marco Liserre,~\IEEEmembership{Fellow,~IEEE}
\thanks{Ziqi Zhou, Sante Pugliese, and Marco Liserre
are with the Chair of Power Electronics,
Kiel University, 24143 Kiel, Germany
(e-mail: ziqi@tf.uni-kiel.de;
sapu@tf.uni-kiel.de;
ml@tf.uni-kiel.de).}
\thanks{Ziqi Zhou, Xiaoqiang Li and Xiaojie Wu are with the
School of Electrical Engineering,
China University of Mining and Technology,
Xuzhou, China
(e-mail: zhouziqi82@163.com; xiaoqiangli@cumt.edu.cn;
xjwu@cumt.edu.cn).}
\thanks{Mahdieh S. Sadabadi is with the
Department of Electrical and Electronic Engineering,
The University of Manchester,
Manchester, United Kingdom
(e-mail: mahdieh.sadabadi@manchester.ac.uk).}
}

\begin{document}
\maketitle

\begin{abstract}
Conventional scalar inertia characterizes only
the low-frequency terminal behavior of
grid-forming converters (GFMs) and cannot fully
predict inertial energy delivery during
short-duration frequency disturbances.
This letter proposes a frequency-response-based
method to assess finite-duration inertial support
using complex terminal power--frequency responses,
such as those provided by Network Frequency
Perturbation (NFP) plots.
For GFMs with nonzero steady-state droop,
a duration-dependent energy-equivalent inertia
is defined after subtracting the ideal
static-droop contribution.
An exact spectral relationship is established,
expressing this metric as a duration-weighted
integral of frequency-dependent inertia,
without requiring knowledge of the internal
controller structure and parameters.
The duration dependence of the metric is further shown to
encode observable oscillation frequencies and damping ratios.
A coefficient-decoupling GFM control with
identical scalar inertia and droop across
different settings is used for verification.
The effects of complex-response noise and finite measurement
bandwidth are evaluated using three paired-panel figures, together
with near-DC regularization and bandwidth-completion methods.
\end{abstract}


\section{Introduction}

Grid-forming (GFM) converters are expected to provide frequency support such as inertia and droop support, yet their terminal power responses are influenced by synchronization and other closed-loop dynamics. Consequently, GFM control designed with identical scalar inertia coefficients may exhibit different transient power and energy delivery during short-duration frequency disturbances~\cite{Eriksson,Han,Jiang}. Conventional time-domain prediction of such responses generally relies on detailed dynamic models, which may not be accessible to system operators due to proprietary control implementations.

Nevertheless, complex terminal power--frequency response data are increasingly available. Apart from direct online-measurement \textcolor{black}{\cite{Shah2023GIST,EPRI2020SSO}}, the recent Great Britain Grid Code requires the submission of a Network Frequency Perturbation (NFP) plot or an agreed equivalent~\cite{NESO}. The NFP plot characterizes the magnitude and phase of the terminal active-power response to frequency perturbations without requiring disclosure of detailed internal control implementations~\cite{Mudalige}. However, how such complex frequency-response data can be directly translated into inertial energy delivery over a prescribed disturbance duration remains insufficiently addressed.

To address this gap, this letter establishes an exact relationship between terminal frequency-dependent inertia and finite-duration inertial energy delivery. A duration-dependent energy-equivalent inertia is introduced to quantify how much inertial support a GFM converter actually delivers over a given time interval. Specifically, it represents the inertia of an ideal system that would deliver the same cumulative inertial energy as the actual GFM converter under the same RoCoF and duration, after removing the static-droop contribution. This metric can be directly calculated from the complex terminal power--frequency response, without requiring knowledge of internal controller structures or parameters.

An analytical time-domain simulation provides a benchmark for validating the proposed frequency-domain formulation. Comparisons among GFM converters designed with identical scalar inertia coefficients reveal different inertial energy delivery over a 50-ms RoCoF interval, demonstrating that identical nominal scalar inertia does not necessarily imply equivalent short-duration inertial support. In addition, the duration dependence of the proposed metric is used to identify oscillatory modes observable in the terminal response.

\section{Motivation of Frequency-Response-Based Finite-Duration Inertia Assessment}

Consider a generic small-signal terminal active-power response
of a GFM converter to grid-frequency
perturbations:
\begin{equation}
F(s)=
-\frac{\Delta P_{\mathrm{GFM}}(s)}
{\omega_n\Delta\omega_g(s)}=\frac{a_\text{n}s^\text{n}+\cdots+a_\text{1}s+a_\text{0}}{b_\text{m}s^\text{m}+\cdots+b_\text{1}s+b_\text{0}},~m\geq n
\label{eq:Fdef}
\end{equation}
where $\omega_n$ and $\omega_g$ denote the nominal
and grid angular frequencies, respectively, and
$P_{\mathrm{GFM}}$ is the injected active power.

Assuming $F(s)$ is \textcolor{black}{asymptotically} stable, proper, real-rational,
and analytic at $s=0$, its Taylor expansion is
\begin{equation}
F(s)=\sum_{k=0}^{\infty}M_k s^k,\qquad
M_k=\frac{F^{(k)}(0)}{k!},
\label{eq:Taylor}
\end{equation}
where $M_k$ denotes the $k$th-order
power--frequency moment.

Retaining the first two terms yields the conventional
scalar droop--inertia approximation~\cite{Eriksson}:
\begin{equation}
\begin{aligned}
F(s)\approx D_{\mathrm{act}}+J_{\mathrm{act}}s,~~
D_{\mathrm{act}}=M_0,~~
J_{\mathrm{act}}=M_1.
\end{aligned}
\label{eq:scalar}
\end{equation}

Consequently, conventional scalar droop $D_{\mathrm{act}}$ and
inertia $J_{\mathrm{act}}$ characterize only the
low-order terminal dynamics ($k\leq 1$), while
higher-order dynamics ($k\geq 2$) are not captured. Accordingly, these scalar coefficients have an asymptotic
time-domain interpretation. Under a sustained
constant RoCoF $r_0$, the terminal power response
of a stable GFM can be predicted
\begin{equation}
\Delta P_{\mathrm{GFM}}(t)
\sim -\omega_n r_0
\left(D_{\mathrm{act}}t+J_{\mathrm{act}}\right),
\quad t\to\infty,
\label{eq:asymptoticPower}
\end{equation}
after the internal transients decay.
This insight also provides a theoretical basis for
conventional scalar inertia
assessments~\cite{Han,Zhou}.

However, this asymptotic equivalence does not guarantee identical finite-duration responses. Even GFMs with identical scalar inertia and droop may exhibit different transient power responses due to their higher-order dynamics. Consequently, the inertial support delivered during the initial stage of a frequency disturbance may differ significantly from that predicted by the conventional scalar-inertia model (see \eqref{eq:scalar}).

Moreover, the explicit transfer function $F(s)$ may not be available to system operators due to confidential controller implementations, making it difficult to analytically determine $D_{\mathrm{act}}$ and $J_{\mathrm{act}}$ through \eqref{eq:Taylor} and \eqref{eq:scalar}. Nevertheless, terminal power--frequency response data may be available through grid-connection requirements or online-measurements \cite{NESO}, providing an alternative means of characterizing GFM dynamics.

These considerations motivate a frequency-response-based approach to quantify the inertial support actually delivered over a finite time interval, without requiring knowledge of the internal controller structure or an explicit transfer-function model.

\section{Terminal Inertia Characterization and
Finite-Duration Inertia Support}

\subsection{Frequency-Dependent Inertia and Droop}

To characterize the terminal dynamics beyond scalar
inertia and droop, frequency-dependent coefficients
are proposed. Within the convergence region
of~\eqref{eq:Taylor}, substituting $s=j\Omega$ yields
\begin{equation}
\begin{aligned}
F(j\Omega)={}&
\sum_{k=0}^{\infty}
(-1)^kM_{2k}\Omega^{2k}
+j\Omega
\sum_{k=0}^{\infty}
(-1)^kM_{2k+1}\Omega^{2k}.
\end{aligned}
\label{eq:Fjw}
\end{equation}

By analogy with the conventional scalar
droop--inertia approximation in~\eqref{eq:scalar},
the real-valued frequency-dependent droop
$D_{\mathrm{eff}}$ and inertia $J_{\mathrm{eff}}$
are defined as
\begin{equation}
\begin{aligned}
D_{\mathrm{eff}}(\Omega)
&=\operatorname{Re}\{F(j\Omega)\},~~
J_{\mathrm{eff}}(\Omega)
=\frac{\operatorname{Im}\{F(j\Omega)\}}{\Omega}.
\end{aligned}
\label{eq:DeffJeff}
\end{equation}
These definitions yield the exact decomposition
\begin{equation}
F(j\Omega)=D_{\mathrm{eff}}(\Omega)
+j\Omega J_{\mathrm{eff}}(\Omega).
\label{eq:Fdecomposition}
\end{equation}
Their low-frequency limits recover the
conventional scalar coefficients:
\begin{equation}
\begin{aligned}
\lim_{\Omega\to0}D_{\mathrm{eff}}(\Omega)
=D_{\mathrm{act}},~~
\lim_{\Omega\to0}J_{\mathrm{eff}}(\Omega)
&=J_{\mathrm{act}}.
\end{aligned}
\label{eq:lowfreq}
\end{equation}
Therefore, scalar droop and inertia can,
in principle, be directly estimated from the low-frequency
terminal response.

Beyond scalar coefficients, $D_{\mathrm{eff}}$
and $J_{\mathrm{eff}}$ retain the high-order (or high frequency) dynamics, providing a more complete
characterization of the GFM power--frequency
response. However, how these characteristics
relate to inertial energy delivery during a
finite-duration frequency disturbance remains
unclear. In particular, whether finite-duration
inertial support can be directly quantified
from the frequency-dependent inertia or droop
characteristics requires further investigation.

The following subsection addresses this question
by establishing an exact relationship between
$J_{\mathrm{eff}}(\Omega)$ and the inertial energy
delivered over a finite-duration constant-RoCoF
interval.

\subsection{Proposed Finite-Duration Inertial Support Assessment}

Let $\Da>0$ and $\Ja=J_0>0$.
To distinguish inertial support from the ideal
static-droop contribution, define
\begin{equation}
H_I(s)\triangleq
\frac{F(s)-\Da}{s},
\qquad H_I(0)=J_0,
\label{eq:HI}
\end{equation}
which indicates the step response of $F_\text{s}$ with the ideal droop contribution excluded. Therefore, $H_I(s)$ only contains the dynamics information of the inertial support.

For a constant angular RoCoF $r_0\ne0$
applied over $0\le t\le T$, the grid-frequency
deviation is $\Delta\omega_g(t)=r_0t$
within this interval.
The corresponding terminal power response is
\begin{equation}
\begin{aligned}
\Delta P_{\mathrm{GFM}}(t)
&=-\wn r_0[\Da t+g_I(t)],\\
g_I(t)
&=\mathcal L^{-1}
\left\{\frac{H_I(s)}{s}\right\},
\quad 0<t<T.
\end{aligned}
\label{eq:power}
\end{equation}
Here, $-\wn r_0\Da t$ represents the ideal
static-droop response, while $g_I(t)$
characterizes the normalized residual
power response associated with transient
inertial support.

After excluding the ideal static-droop contribution,
the cumulative inertial energy exchange is
\begin{equation}
E_I(T)=-\wn r_0\int_0^T g_I(t)\,\dd t
=-\wn r_0 T\Js(T),
\label{eq:EI}
\end{equation}
where $\Js(T)$ is the proposed \textit{equivalent scalar inertia}
producing the same cumulative energy exchange
over the prescribed RoCoF interval. Accordingly,
\begin{equation}
\Js(T)=\frac{1}{T}\int_0^Tg_I(t)\,\dd t.
\label{eq:Js}
\end{equation}

It therefore characterizes the average
inertial support within given duration rather than the
instantaneous power response.

To establish the relationship between
$\Js(T)$ and $\Je(\Omega)$, consider a
rectangular RoCoF input
$w_T(t)=u(t)-u(t-T)$, whose Fourier
transform is
$W_T(j\Omega)=T e^{-j\Omega T/2}
\sinc(\Omega T/2)$, \textcolor{black}{where $\sinc(x)=\sin(x)/x$}.
Let $h_I(t)=\mathcal L^{-1}\{H_I(s)\}$.
Then \eqref{eq:power} can be rewritten as $g_I(t)
=(h_I*w_T)(t)=y_T(t),~0<t<T.$

Based on this, \eqref{eq:Js} can be formulated as
\begin{equation}
\begin{aligned}
\Js(T)
=\frac{1}{T}\int_0^Ty_T(t)\,\dd t=\frac{1}{T}\int_{-\infty}^{\infty}y_T(t)w_T(t)\,\dd t
\end{aligned}
\label{eq:Js_yt}
\end{equation}

Applying Parseval's identity (Eq. 2.35 in \cite{MIT})
to~\eqref{eq:Js_yt} yields
\begin{equation}
\Js(T)=\frac{1}{2\pi T}
\int_{-\infty}^{\infty}
H_I(j\Omega)
|W_T(j\Omega)|^2\,\dd\Omega.
\label{eq:Parseval}
\end{equation}

Since $F(s)$ is real-rational, $H_I(s)$
satisfies the conjugate symmetry
$H_I(-j\Omega)=H_I^*(j\Omega)$.
Together with
$\operatorname{Re}\{H_I(j\Omega)\}
=J_{\mathrm{eff}}(\Omega)$,
this yields
\begin{equation}
\begin{aligned}
\Js(T)
&=\int_0^\infty
\mathcal W_T(\Omega)\Je(\Omega)\,\dd\Omega,\\
\mathcal W_T(\Omega)
&=\frac{T}{\pi}\sinc^2
\left(\frac{\Omega T}{2}\right).
\end{aligned}
\label{eq:spectrum}
\end{equation}
Here, $\mathcal W_T(\Omega)$ is a
nonnegative, normalized weighting function.

Equation~\eqref{eq:spectrum} establishes
an exact relationship between the terminal
inertia spectrum and finite-duration
inertial energy delivery.
It demonstrates that $\Js(T)$ can be
determined solely from $\Je(\Omega)$
for any prescribed RoCoF duration,
without requiring knowledge of the
internal GFM control structure.

Furthermore, for a stable $H_I$,
\begin{equation}
\lim_{T\to\infty}\Js(T)
=\lim_{\Omega\to0}\Je(\Omega)
=J_{\mathrm{act}}.
\label{eq:JsLimit}
\end{equation}
Therefore, conventional scalar inertia
represents the long-duration limit
of the proposed energy-equivalent inertia.
Beyond the conventional scalar inertia, $\Js(T)$ generalizes scalar
inertia by characterizing inertial energy
delivery over arbitrary finite durations.

This enables system operators to assess
and compare the duration-dependent
inertial support of different GFMs
directly from their terminal frequency
responses.
For a prescribed duration $T$, a larger
$J_{\mathrm{sup}}(T)$ indicates greater
cumulative inertial energy delivery
under the same constant RoCoF duration.

\subsection{Oscillation Damping Identification Through $J_{\mathrm{sup}}(T)$}
\label{sec:osc_damping}

Although $J_{\mathrm{sup}}(T)$ is introduced as an
energy-equivalent inertia, its dependence on the
observation duration also encodes the terminal
oscillation dynamics. Define $S(T)=T J_{\mathrm{sup}}(T)$.
From~\eqref{eq:Js}, and using
$g_I(t)=\int_0^t h_I(\tau)\,\mathrm{d}\tau$,
we obtain
\begin{equation}
\begin{aligned}
\frac{\mathrm{d}S(T)}{\mathrm{d}T}&=g_I(T),\\
q_I(T)\triangleq\frac{\mathrm{d}^2 S(T)}{\mathrm{d}T^2}
&=h_I(T).
\end{aligned}
\label{eq:osc_recovery}
\end{equation}
Therefore, knowledge of $J_{\mathrm{sup}}(T)$ over a
range of durations is sufficient, in principle, to
recover the impulse response of $H_I(s)$.
The temporal averaging involved in the inertia
metric smooths the oscillations but does not
eliminate their underlying modal information.

For clarity, assume that $H_I(s)$ has distinct,
nonzero poles $p_k$ with residues $c_k$. Since
$H_I(s)$ is strictly proper, its modal expansion
and~\eqref{eq:osc_recovery} yield
\begin{equation}
\begin{aligned}
q_I(T)&=\sum_{k=1}^{N}c_k e^{p_kT},\\
J_{\mathrm{sup}}(T)&=J_{\mathrm{act}}+
\frac{1}{T}\sum_{k=1}^{N}
\frac{c_k}{p_k^2}\big(e^{p_kT}-1\big).
\end{aligned}
\label{eq:osc_modal}
\end{equation}
In particular, a complex-conjugate mode
$p_m=-\sigma_m\pm j\omega_{d,m}$ contributes
\begin{equation}
q_{I,m}(T)=2|c_m|e^{-\sigma_mT}
\cos(\omega_{d,m}T+\phi_m),
\label{eq:osc_pair}
\end{equation}
where $\phi_m=\arg(c_m)$. Thus, the oscillation
frequency and damping ratio can be identified as
\begin{equation}
\begin{aligned}
f_{d,m}&=\frac{\omega_{d,m}}{2\pi},\\
\zeta_m&=\frac{\sigma_m}
{\sqrt{\sigma_m^2+\omega_{d,m}^2}}.
\end{aligned}
\label{eq:osc_zeta}
\end{equation}
Consequently, the duration dependence of
$J_{\mathrm{sup}}(T)$ provides an additional means
of identifying oscillatory modes from the complex
terminal frequency response through~\eqref{eq:spectrum},
without requiring the internal controller model.

\section{Analytical Benchmark and Verification}

\subsection{Coefficient-Decoupling GFM With Fixed Scalar
Inertia and Droop Design}

Consider the reduced coefficient-decoupling GFM
model derived from~\cite{Zhou}:
\begin{equation}
F(s)=
\frac{K_s(D_ds+K_iD_a)}
{D_ds^2+(K_iD_a+K_s)s+K_sK_i}.
\label{eq:model}
\end{equation}
According to \eqref{eq:lowfreq}, the corresponding scalar droop and inertia are
\begin{equation}
\begin{aligned}
D_{\mathrm{act}}&=D_a,~~
J_{\mathrm{act}}
=\frac{D_d-D_a}{K_i}
-\frac{D_a^2}{K_s}.
\end{aligned}
\label{eq:Jact}
\end{equation}
To maintain identical scalar inertia $J_0$
and droop $D_a$ across different control
parameters, $D_d$ is selected as
\begin{equation}
D_d=D_a+K_i
\left(J_0+\frac{D_a^2}{K_s}\right).
\label{eq:Dd}
\end{equation}
Substituting~\eqref{eq:model} into~\eqref{eq:HI}
yields
\begin{equation}
H_I(s)=
\frac{K_sK_iJ_0-D_aD_ds}
{D_ds^2+(K_iD_a+K_s)s+K_sK_i}.
\label{eq:HIreduced}
\end{equation}

The parameters are $f_{\mathrm n}=50$~Hz,
$J_0=1$,
$K_s=11251.8907$,
$D_a=5$, and
$K_i\in\{40,85,160,2400\}$.
By constructing $D_\text{d}$ according to \eqref{eq:Dd}, all cases have identical
$D_{\mathrm{act}}=5$ and
$J_{\mathrm{act}}=J_0$, but with varying $J_\text{eff}$.

\begin{figure}[!t]
\centering
\includegraphics[width=\columnwidth]
{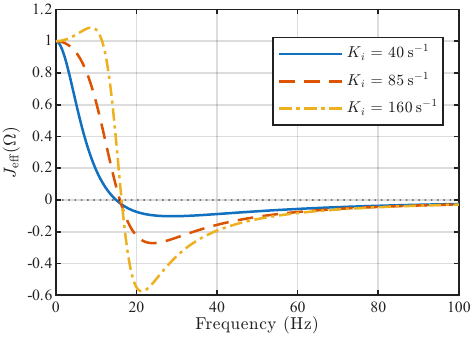}
\caption{Frequency-dependent inertia for
three GFM cases with identical scalar
inertia and droop.}
\label{fig:response}
\end{figure}

Fig.~\ref{fig:response} compares their
$J_\text{eff}$.
Although all cases share the same
low-frequency inertia limit, their
frequency-dependent characteristics
differ significantly.
This indicates that identical scalar inertia
does not necessarily imply identical
transient inertial support.

\subsection{Time-Domain Analytical Benchmark and Frequency-Domain Validation}

When the terminal transfer function is known,
the normalized inertial power response
$g_I(t)$ can be obtained directly through
the inverse Laplace transformation of
$H_I(s)/s$.
Substituting this response into~\eqref{eq:Js}
provides the exact duration-dependent
energy-equivalent inertia for the
considered linear model.
This time-domain solution serves as an
independent analytical benchmark for
validating the proposed frequency-domain
assessment method.

While the analytical approach requires
an explicit transfer-function model,
the proposed method evaluates
$J_{\mathrm{sup}}(T)$ directly from the
complex terminal frequency response
using~\eqref{eq:spectrum}.
Both approaches are mathematically equivalent
when the complete frequency response
is available.

\begin{figure}[!t]
\centering
\includegraphics[width=\columnwidth]
{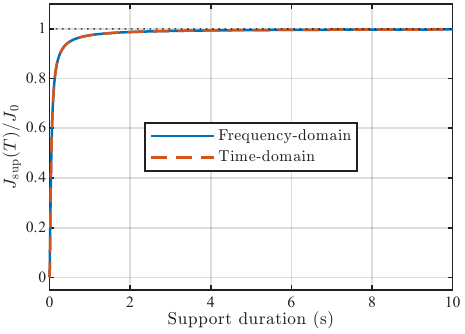}
\caption{Duration-dependent energy-equivalent
inertia computed through both time-domain and frequency-domain approaches.}
\label{fig:TD_FD_Jsup}
\end{figure}

To validate the identical $J_\text{sup}(T)$ curves between the time-domain (see \eqref{eq:Js}) and the proposed frequency-domain approaches (see \eqref{eq:spectrum}), Fig. \ref{fig:TD_FD_Jsup} shows the dynamics of $J_\text{sup}(T)$ with $K_i=40$. For the proposed method, the complex terminal frequency response
$F(j\Omega)$ is sampled at 1001 logarithmically
spaced frequencies between 0.0001 and 100~Hz.

It can be seen the both approaches exhibit identical trajectories of $J_\text{sup}(T)$ within 10~s duration, and finally converge to $J_0$. These results validate the effectiveness of the proposed frequency-domain method.

\subsection{Finite-Duration Inertial Support Evaluation}

To validate the proposed method,
the complex terminal frequency response
$F(j\Omega)$ is sampled at 601 logarithmically
spaced frequencies between 0.01 and 50~Hz.
The frequency-dependent inertia
$J_{\mathrm{eff}}(\Omega)$ is obtained
from~\eqref{eq:DeffJeff}, and
$J_{\mathrm{sup}}(T)$ is calculated
through \eqref{eq:spectrum}.

\begin{figure}[!t]
\centering
\includegraphics[width=\columnwidth]
{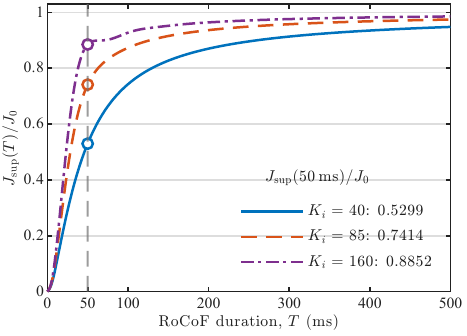}
\caption{Duration-dependent energy-equivalent
inertia for three GFM cases with identical
scalar inertia and droop.}
\label{fig:delivery}
\end{figure}

Fig.~\ref{fig:delivery} illustrates
the duration-dependent energy-equivalent
inertia obtained from the analytical
benchmark.
Despite identical scalar inertia,
the three cases exhibit considerably
different inertial energy delivery
over finite durations.
For instance, at $T=50$~ms,
$J_{\mathrm{sup}}(T)/J_0$ increases
from 0.5297 to 0.8851 as $K_i$
increases from 40 to 160.

Table~\ref{tab:results} compares the
frequency-domain estimates with the
analytical benchmark for a prescribed
50-ms RoCoF duration.
The maximum relative error is below
0.23\%, demonstrating good agreement
between the two approaches for the
investigated frequency range and
sampling resolution.

In addition, as the duration increases, all curves
gradually approach the conventional
scalar inertia $J_{\mathrm{act}}$,
consistent with~\eqref{eq:JsLimit}.
These results confirm that scalar inertia
alone cannot fully characterize inertial
support over a prescribed finite duration.

\begin{table}[!t]
\centering
\caption{Comparison of 50-ms Energy-Equivalent Inertia}
\label{tab:results}
\footnotesize
\setlength{\tabcolsep}{5pt}
\begin{tabular}{@{}ccccc@{}}
\toprule
\multirow{2}{*}{$K_i$} &
\multirow{2}{*}{$\zeta$} &
\multicolumn{2}{c}{$J_{\mathrm{sup}}(50\,\mathrm{ms})/J_0$} &
\multirow{2}{*}{Error (\%)} \\
\cmidrule(lr){3-4}
& & Benchmark & Proposed & \\
\midrule
40  & 1.271 & 0.5297 & 0.5309 & 0.228 \\
85  & 0.629 & 0.7412 & 0.7428 & 0.211 \\
160 & 0.349 & 0.8851 & 0.8868 & 0.192 \\
\bottomrule
\end{tabular}
\end{table}

\subsection{Finite-Duration Inertial Energy}

To further illustrate the physical meaning
of $J_{\mathrm{sup}}(T)$, a constant RoCoF
of $r_0=-2\pi$~rad/s$^2$ is applied
over a 50-ms interval. In this condition, an ideal scalar inertia behavior predicts an inertia energy of 98.65 J.

\begin{figure}[!t]
\centering
\includegraphics[width=\columnwidth]{
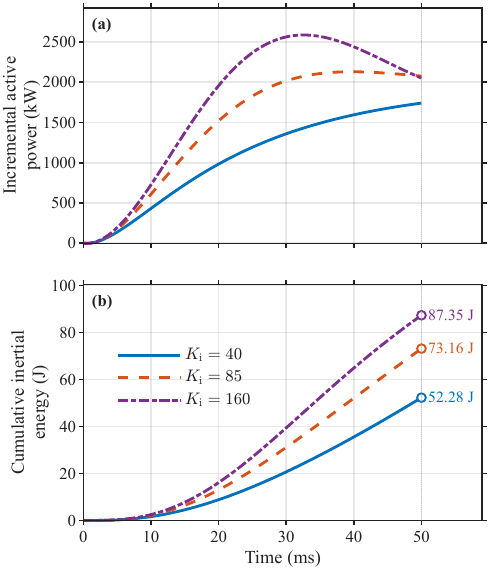}
\caption{Transient responses of three
GFM cases within a 50-ms constant-RoCoF
interval: (a) terminal active-power
responses and the ideal static-droop
contribution; (b) cumulative inertial
energy after subtracting the ideal
static-droop contribution.}
\label{fig:power}
\end{figure}

Fig.~\ref{fig:power}(a) compares the
corresponding terminal active-power
responses and the ideal static-droop
contribution.
Despite identical scalar coefficients,
the three cases exhibit distinct
transient power responses.

The cumulative inertial energy within
the RoCoF interval is defined as
\begin{equation}
E_I(t)=
\int_0^t
\big[
\Delta P_{\mathrm{GFM}}(\tau)
+\omega_nD_{\mathrm{act}}r_0\tau
\big]\mathrm{d}\tau,
\label{eq:inertialEnergy}
\end{equation}
where $0\leq t\leq T_r$ and
$T_r=50$~ms.
According to~\eqref{eq:Js}, the cumulative
inertial energy at the end of the interval
satisfies
\begin{equation}
E_I(T_r)=
-\omega_n r_0T_rJ_{\mathrm{sup}}(T_r).
\label{eq:energyEquivalence}
\end{equation}

As shown in Fig.~\ref{fig:power}(b),
the three cases deliver 52.28, 73.16,
and 87.35~J of inertial energy within
the 50-ms interval, corresponding to
52.97\%, 74.12\%, and 88.51\% of the
98.70~J predicted by the conventional
scalar-inertia model, respectively. These ratios match with the $J_{\mathrm{sup}}(50\,\mathrm{ms})/J_0$ in Table~\ref{tab:results}, validating the effectiveness of the proposed method.

\subsection{Oscillation Damping Identification Verification}
\label{subsec:osc_validation}

To verify the proposed modal interpretation, the
benchmark in~\eqref{eq:model} is evaluated for
$K_i\in\{40,85,160,2400\}$ while maintaining
$J_{\mathrm{act}}=J_0=1$ and $D_{\mathrm{act}}=5$.
Its reference modal parameters follow directly
from the denominator in~\eqref{eq:model}:
\begin{equation}
\begin{aligned}
\sigma&=\frac{K_iD_a+K_s}{2D_d},\\
\omega_d&=\sqrt{\frac{K_sK_i}{D_d}-\sigma^2},\qquad
\zeta=\frac{\sigma}{\sqrt{K_sK_i/D_d}},
\end{aligned}
\label{eq:osc_benchmark}
\end{equation}
where $\omega_d$ is real for the underdamped cases.
At $K_i=40$, $\zeta=1.271>1$, and the two poles
are real; the other settings have oscillatory modes.

For identification, $F(j\Omega)$ is sampled at
1501 logarithmically spaced frequencies from
$10^{-4}$ to 200~Hz. The resulting
$J_{\mathrm{eff}}(\Omega)$ is substituted
into~\eqref{eq:spectrum} to calculate
$J_{\mathrm{sup}}(T)$ at 0.5-ms duration increments.
The response $q_I(T)$ is then reconstructed using
\eqref{eq:osc_recovery} and fitted to a decaying
sinusoid to identify $f_d$ and $\zeta$. The
analytical poles are used only as an independent
reference, not during identification.

\begin{figure}[!t]
\centering
\includegraphics[width=\columnwidth]
{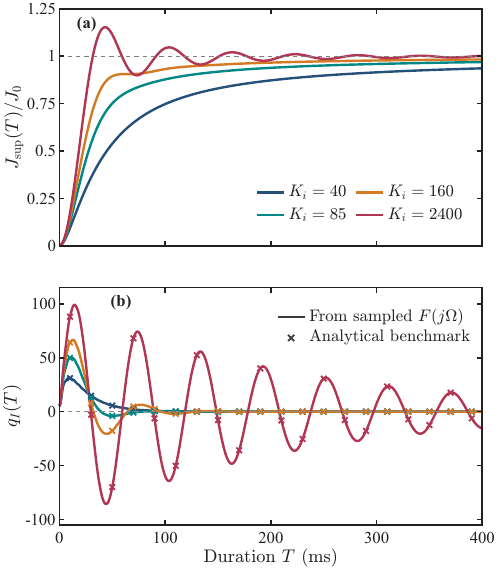}
\caption{Identification of terminal oscillation modes
from $J_{\mathrm{sup}}(T)$: (a) duration-dependent
energy-equivalent inertia; (b) reconstructed
$q_I(T)=\mathrm{d}^2[TJ_{\mathrm{sup}}(T)]/\mathrm{d}T^2$.
Solid curves are obtained from sampled terminal
frequency responses; crosses indicate the
analytical impulse-response benchmark.}
\label{fig:osc_identification}
\end{figure}

As shown in Fig.~\ref{fig:osc_identification},
the weakly damped case ($K_i=2400$) exhibits
pronounced oscillations in $J_{\mathrm{sup}}(T)$,
whereas more strongly damped modes can be obscured
by duration averaging. Nevertheless, these
oscillations are recovered in $q_I(T)$, which
agrees with the independent impulse-response
benchmark. Table~\ref{tab:osc_modes} further shows
that the estimated modal frequencies agree with
the analytical values within 0.22\%, and the
estimated damping ratios within 0.05\% in relative
terms for the investigated underdamped cases.

\begin{table}[!t]
\centering
\caption{Modal Identification From $J_{\mathrm{sup}}(T)$}
\label{tab:osc_modes}
\footnotesize
\setlength{\tabcolsep}{3.2pt}
\begin{tabular}{@{}ccccc@{}}
\toprule
\multirow{2}{*}{$K_i$} &
\multicolumn{2}{c}{$f_d$ (Hz)} &
\multicolumn{2}{c}{$\zeta$} \\
\cmidrule(lr){2-3}\cmidrule(lr){4-5}
& Pole & From $J_{\mathrm{sup}}$ & Pole & From $J_{\mathrm{sup}}$ \\
\midrule
40   & --     & --     & 1.2711 & -- \\
85   & 12.746 & 12.718 & 0.6286 & 0.6285 \\
160  & 15.561 & 15.552 & 0.3493 & 0.3494 \\
2400 & 16.829 & 16.829 & 0.04557 & 0.04558 \\
\bottomrule
\end{tabular}
\end{table}

The higher identification bandwidth, compared with
the 50-Hz finite-energy assessment, is used to
limit numerical-differentiation errors. For noisy
or bandwidth-limited measurements, direct fitting
of~\eqref{eq:osc_modal} is preferable. The
identified quantities characterize observable
terminal modes, rather than all internal system
modes.

% ============================================================
% SECTION V: MEASUREMENT ROBUSTNESS
% ============================================================

\section{Robustness to Imperfect Frequency-Response Measurements}
\label{sec:measurement_robustness}

The identity in~\eqref{eq:spectrum} is exact for a complete,
noise-free frequency response. In practice, terminal measurements
are acquired at finitely many frequencies and may contain measurement
noise or have limited measurement bandwidth. These limitations are particularly relevant
to the finite-duration metric because its weighting kernel depends
on the prescribed support duration. Introducing frequency in hertz,
$f=\Omega/(2\pi)$, gives the equivalent representation
\begin{equation}
\begin{aligned}
\Js(T)&=\int_0^{\infty}K_T(f)\Je(2\pi f)\,\dd f,\\
K_T(f)&=2T\sinc^2(\pi fT),\quad
\int_0^{\infty}K_T(f)\,\dd f=1.
\end{aligned}
\label{eq:robust_kernel}
\end{equation}
Here, $\sinc(x)=\sin(x)/x$ as in~\eqref{eq:spectrum}.
The short-duration kernel spreads over a broad frequency range,
whereas its long-duration counterpart concentrates near zero
frequency. Consequently, the impacts of noise and finite bandwidth
depend on the prescribed duration $T$.

The illustrations below use the reduced second-order terminal
model in~\eqref{eq:model} with $K_i=2400$, $J_0=1$,
$D_a=5$, and the previously specified $K_s$ and $D_d$.
This setting features a weakly damped mode near 16.8~Hz.
Unless stated otherwise, the exact modal response
in~\eqref{eq:osc_modal} is an independent reference,
not an input to the measurement-data corrections.
Every robustness figure contains two vertically stacked
panels: (a) above (b).

\subsection{Measurement Noise and Near-DC Regularization}
\label{subsec:noise_robustness}

Suppose that the measured complex terminal response at
$f_k>0$ is
\begin{equation}
\widehat F_k=F(j2\pi f_k)+\eta_k,\qquad
\widehat J_{\mathrm{eff},k}=\Je(2\pi f_k)
+\frac{\operatorname{Im}\{\eta_k\}}{2\pi f_k},
\label{eq:robust_noisy_je}
\end{equation}
where $\eta_k$ is a complex measurement error. Even when the
imaginary-part error is small and has constant variance, its
contribution to the inferred inertia is amplified by $1/f_k$.
For trapezoidal integration, let $q_k$ denote the quadrature
weights in hertz, and define
$a_k(T)=q_kK_T(f_k)/(2\pi f_k)$.
Then the noise-induced error obeys
\begin{equation}
\begin{aligned}
\varepsilon_n(T)&=\widehat\Js(T)-\Js^{\mathrm{sampled}}(T)
=\sum_k a_k(T)\operatorname{Im}\{\eta_k\},\\
\operatorname{var}[\varepsilon_n(T)]
&=\boldsymbol a(T)^{\mathsf T}\boldsymbol\Sigma_I
\boldsymbol a(T),
\end{aligned}
\label{eq:robust_noise_variance}
\end{equation}
where $\boldsymbol\Sigma_I$ is the covariance matrix of the
imaginary measurement errors. For independent samples, the
variance reduces to $\sum_k a_k^2(T)\sigma_{I,k}^2$.
Equation~\eqref{eq:robust_noise_variance} also shows why an
extremely small first frequency combined with a relatively
large trapezoidal cell can dominate the uncertainty. Correlated
noise must be handled using the full covariance matrix.

For a real-coefficient linear time-invariant terminal response,
$\operatorname{Im}\{F(j2\pi f)\}$ is odd and vanishes at $f=0$.
Instead of smoothing $\widehat J_{\mathrm{eff}}$ after division
by $2\pi f$, we therefore fit the measured imaginary response
on a near-DC frequency interval using
\begin{equation}
\begin{aligned}
\operatorname{Im}\{\widehat F(j2\pi f)\}
&\simeq \sum_{\ell=0}^{L}b_{2\ell+1}f^{2\ell+1},\\
\widehat J_{\mathrm{act}}&=\frac{b_1}{2\pi}.
\end{aligned}
\label{eq:robust_odd_fit}
\end{equation}
The fitted polynomial replaces only the designated low-frequency
samples. The fit must remain below relevant terminal resonances:
a wide or overly flexible fitting interval can remove actual
small-signal dynamics rather than measurement noise. Averaging
repeated frequency-response measurements and using weighted
least squares provide further variance reduction when the
measurement uncertainty is known.

\begin{figure}[!t]
\centering
\includegraphics[width=\columnwidth]{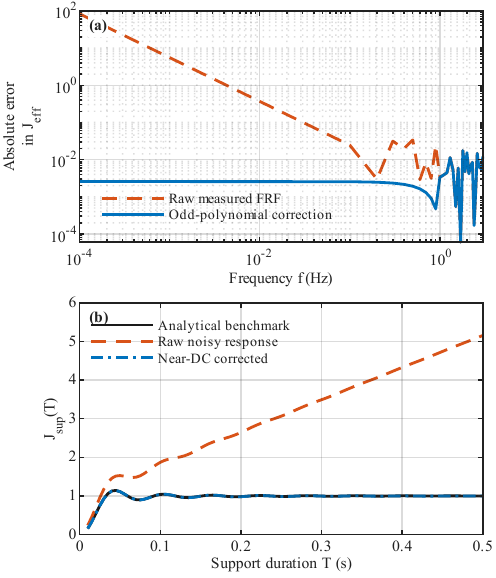}
\caption{Complex frequency-response measurement noise at 1\% relative
RMS: (a) error amplification in $J_{\mathrm{eff}}$ near DC and
its suppression by an odd polynomial fit of $\operatorname{Im}\{F\}$;
(b) corresponding direct and corrected $J_{\mathrm{sup}}(T)$,
compared with the analytical benchmark. (FRF denotes Frequency Response Function)}
\label{fig:robust_noise}
\end{figure}

In Fig.~\ref{fig:robust_noise}(a), division by $2\pi f$
turns a small imaginary-response error into a large near-DC
inertia error; the odd-polynomial fit suppresses the resulting
spikes. The same single noisy realization is propagated to
$J_{\mathrm{sup}}(T)$ in Fig.~\ref{fig:robust_noise}(b),
where the corrected estimate more closely follows the
independent analytical reference. The correction requires
only measured complex-response samples, not the true poles
or scalar inertia.

\subsection{Finite Measurement Bandwidth}
\label{subsec:bandwidth_robustness}

If measurements are available only for $f\in[f_L,f_H]$, the
uncompensated band-limited assessment is
$\Js^{[f_L,f_H]}(T)=\int_{f_L}^{f_H}
K_T(f)\Je(2\pi f)\,\dd f$. Its truncation bias is exactly
\begin{equation}
\begin{aligned}
\varepsilon_B(T)
={}&-\int_0^{f_L}K_T(f)\Je(2\pi f)\,\dd f\\
&-\int_{f_H}^{\infty}K_T(f)\Je(2\pi f)\,\dd f.
\end{aligned}
\label{eq:robust_band_bias}
\end{equation}
If $|\Je(2\pi f)|\le M_L$ below $f_L$ and
$|\Je(2\pi f)|\le M_H$ above $f_H$, then
$K_T(f)\le2T$ and
$K_T(f)\le2/(\pi^2Tf^2)$ imply the conservative bound
\begin{equation}
|\varepsilon_B(T)|\le
2M_LT f_L+\frac{2M_H}{\pi^2T f_H}.
\label{eq:robust_band_bound}
\end{equation}
The two terms exhibit opposite duration dependences:
a finite lower cutoff primarily compromises sufficiently long
durations, whereas an insufficient upper cutoff is most
restrictive at short durations. This estimate is conservative
and requires justified bounds on the unmeasured spectrum.

A measurement-only correction for the lower cutoff follows
from the odd fit in~\eqref{eq:robust_odd_fit}, which extends
$\widehat J_{\mathrm{eff}}$ continuously to $f=0$. At sufficiently
high frequencies, a strictly proper real-rational response
frequently yields $\Je(2\pi f)=O(f^{-2})$; hence, a data-fitted
$A/f^2$ continuation can estimate the missing upper tail,
provided that the measured end band has already reached
this asymptotic regime. Such extrapolation is unreliable
when $f_H$ is below, or close to, an unresolved resonance.
Without a valid model or prior spectral bounds, the unmeasured
contribution cannot be uniquely reconstructed, and an
uncertainty interval should be reported instead.

\begin{figure}[!t]
\centering
\includegraphics[width=\columnwidth]{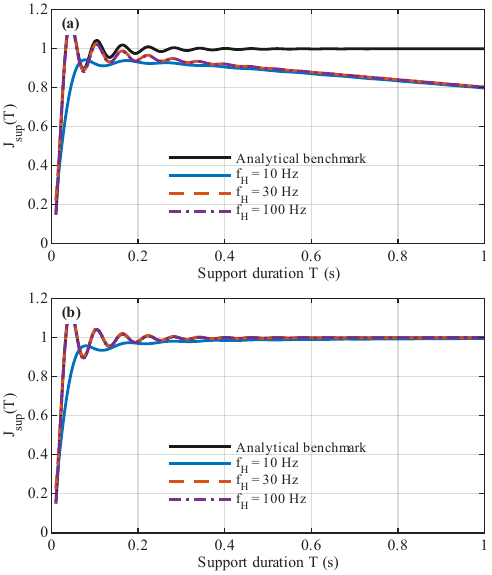}
\caption{Influence of measured bandwidth $[0.1,f_H]$~Hz on
$J_{\mathrm{sup}}(T)$ for $f_H=10$, 30, and 100~Hz:
(a) direct band-limited integration and (b) low-frequency
completion, additionally applying an asymptotic upper-tail
continuation only for $f_H=100$~Hz.}
\label{fig:robust_bandwidth}
\end{figure}

\begin{figure}[!t]
\centering
\includegraphics[width=\columnwidth]{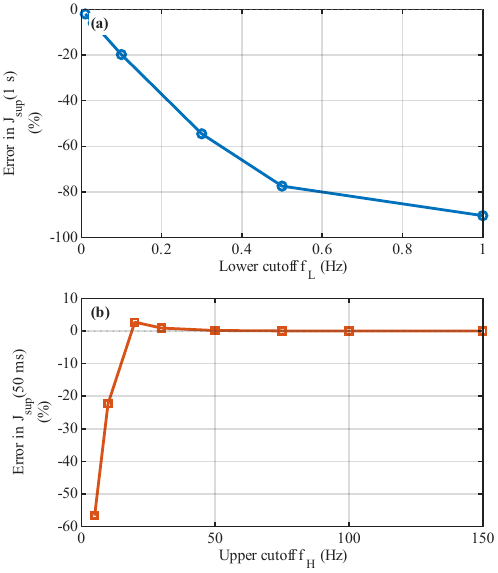}
\caption{Separation of low- and high-frequency truncation:
(a) relative bias at $T=1$~s versus lower cutoff $f_L$
with $f_H=150$~Hz and (b) relative bias at $T=50$~ms
versus $f_H$ with a reconstructed DC sample. }
\label{fig:robust_cutoffs}
\end{figure}

Fig.~\ref{fig:robust_bandwidth}(a) compares direct integration
of the measured $[0.1,f_H]$-Hz bands, while
Fig.~\ref{fig:robust_bandwidth}(b) adds a data-fitted
near-DC continuation. An $A/f^2$ high-frequency tail is
included only for the 100-Hz case; no upper-tail recovery
is claimed for the 10- or 30-Hz bands.

The complementary cutoff sweep in
Fig.~\ref{fig:robust_cutoffs}(a) varies $f_L$ while keeping
$f_H=150$~Hz and evaluates the relative error at $T=1$~s.
Fig.~\ref{fig:robust_cutoffs}(b) varies $f_H$ with the
DC limit included and assesses the error at $T=50$~ms.
Together, the two figures illustrate why the measured
bandwidth must be selected for the duration of interest,
not solely for a nominal resonance frequency.
The signed errors need not vary monotonically because
$J_{\mathrm{eff}}$ may change sign across a resonance.

\subsection{Numerical Illustration and Measurement Guidelines}
\label{subsec:robust_verification}

For all three robustness figures, the same benchmark and spectral
weighting are used. The complex multiplicative noise is
injected into $F(j2\pi f)$ rather than directly into $\Js$.
The data-driven near-DC fit uses the 0.1--2-Hz interval and
replaces samples below 1~Hz. The band-limited experiments
vary the upper and lower frequency cutoffs separately, so the
contribution of each truncation mechanism can be assessed.

\begin{table}[!t]
\centering
\caption{Illustrative Sensitivity of $J_{\mathrm{sup}}(T)$ to Limited Measurement Bandwidth ($K_i=2400$)}
\label{tab:robust_numeric}
\footnotesize
\setlength{\tabcolsep}{3.4pt}
\begin{tabular}{@{}lcc@{}}
\toprule
Frequency-response data & $T=50$ ms & $T=1$ s \\
\midrule
Analytical reference & 1.1090 & 0.9988 \\
0.1--100 Hz, truncated & 1.0992 & 0.8020 \\
0.1--100 Hz, completed & 1.1090 & 0.9998 \\
\bottomrule
\end{tabular}
\end{table}

Table~\ref{tab:robust_numeric} reports an independent
noise-free numerical cross-check. For the 0.1--100~Hz band,
``completed'' denotes near-DC odd extrapolation and a fitted
$A/f^2$ high-frequency tail.
Taken together, the three paired-panel figures illustrate
near-DC noise amplification and its regularization
(Fig.~\ref{fig:robust_noise}), together with duration-dependent
bandwidth bias and cutoff sensitivity
(Figs.~\ref{fig:robust_bandwidth} and~\ref{fig:robust_cutoffs}).
In a practical assessment, the measured frequency band,
sampling grid, noise assumptions, fitting interval, and
extrapolation procedure should be disclosed together with a
duration-dependent uncertainty estimate.

\section{Conclusion}

This letter presents a terminal
frequency-response-based method for
assessing finite-duration inertial energy
support of GFMs with nonzero steady-state
droop.
An exact spectral relationship is established
between duration-dependent energy-equivalent
inertia and the terminal frequency-dependent
inertia, without requiring internal
controller models.
Analytical and frequency-domain assessments
agree within 0.23\% for the investigated
50-Hz measurement bandwidth.
Despite identical scalar inertia and droop,
the three GFM cases deliver only 52.97\%,
74.12\%, and 88.51\% of the ideal
scalar-inertia energy within 50~ms.
These findings demonstrate the limitations
of scalar inertia in characterizing
short-duration energy delivery and provide
a basis for assessing inertial support
from terminal frequency-response data.
Furthermore, the duration dependence of
$J_{\mathrm{sup}}(T)$ preserves information about
observable terminal oscillation modes, enabling
the estimation of their frequencies and damping ratios.
The measurement-robustness analysis further quantifies
noise amplification and duration-dependent bandwidth bias.
These results motivate near-DC regularization and
duration-aware measurement and uncertainty reporting.


\begin{thebibliography}{9}
\bibitem{Eriksson} R.~Eriksson, N.~Modig, and K.~Elkington, ``Synthetic inertia versus fast frequency response: A definition,'' \emph{IET Renewable Power Generation}, vol.~12, no.~5, pp.~507--514, 2018.
\bibitem{Han} G.~Han \emph{et al.}, ``A simple estimation method of grid-forming inverter inertia based on rate of change of frequency,'' \emph{IET Power Electronics}, vol.~17, no.~5, pp.~591--602, 2024.
\bibitem{Jiang} S.~Jiang, Y.~Zhu, T.~Xu, X.~Wang, and G.~Konstantinou, ``Frequency domain inertia design of grid-forming converters,'' \emph{IEEE Transactions on Power Electronics}, vol.~40, no.~7, pp.~8886--8898, 2025.
\bibitem{Shah2023GIST} S.~Shah, P.~Koralewicz, E.~Mendiola, V.~Gevorgian, R.~Wallen, and W.~Yan, ``Grid Impedance Scan Tool (GIST): Software for Stability Analysis of IBR Power Systems,'' presented at the EPRI Sub-Synchronous Oscillations (SSO) Workshop, 2023.
\bibitem{EPRI2020SSO} Electric Power Research Institute, ``Sub-Synchronous Oscillation Screening Through Impedance-Based Frequency Scans: Impact of Converter Control Designs and System Strength,'' Tech. Rep.~3002020209, 2020.
\bibitem{NESO} National Energy System Operator, ``The grid code,'' Tech. Rep., May 2026, Issue~6, Revision~38, Planning Code PC.A.5.8.1(i).
\bibitem{Mudalige} A.~Mudalige, H.~Wu, X.~Wang, M.~Langwasser, and M.~Liserre, ``Influence of the reactive power control loop on network frequency perturbation plot,'' in \emph{Proc. IEEE PEDG}, pp.~814--820, 2023.
\bibitem{Zhou} Z.~Zhou, S.~Pugliese, M.~Langwasser, and M.~Liserre, ``Subsynchronous damping by battery storage system in grid-forming control,'' \emph{IEEE Transactions on Power Electronics}, vol.~39, no.~4, pp.~4173--4186, 2024.
\bibitem{MIT} A.~V.~Oppenheim and G.~C.~Verghese, \emph{Signals, Systems and Inference}. Cambridge, MA, USA: Massachusetts Institute of Technology, 2010, ch.~2.
\end{thebibliography}
\end{document}